\documentclass[runningheads]{llncs}
\usepackage{graphicx}
\usepackage{wrapfig}
\usepackage{subcaption}
\usepackage{float}
\usepackage{amsmath}

\begin{document}
\title{Modelling service-oriented systems and cloud services with \normalfont \textsc{Heraklit}}
%
%

\author{Peter Fettke\inst{1,2}\orcidID{0000-0002-0624-4431} \and
Wolfgang Reisig\inst{3}\orcidID{0000-0002-7026-2810}}
\authorrunning{P. Fettke, W. Reisig}
%

\institute{German Research Center for Artificial Intelligence (DFKI), Saarbr\"ucken, Germany \\
\email{peter.fettke@dfki.de}\\ \and
Saarland University, Saarbr\"ucken, Germany \\ \and
Humboldt-Universität zu Berlin, Berlin, Germany \\  
\email{reisig@informatik.hu-berlin.de}}

\maketitle              
\begin{abstract}
Modern and next generation digital infrastructures are technically based on service oriented structures, cloud services, and other architectures that compose large systems from smaller subsystems. The composition of subsystems is particularly challenging, as the subsystems themselves may be represented in different languages, modelling methods, etc. It is quite challenging to precisely conceive, understand, and represent this kind of technology, in particular for a given level of abstraction. To capture refinement and abstraction principles, various forms of ``technology stacks'' and other semi-formal or natural language based on presentations have been suggested. Generally, useful concepts to compose such systems in a systematic way are even more rare. \textsc{Heraklit} provides means, principles, and unifying techniques to model and to analyze digital infrastructures.
\textsc{Heraklit} integrates composition and hierarchies of subsystems, concrete and abstract data structures, as well as descriptions of behaviour. A distinguished set of means supports the modeler to express their ideas. The modeller is free to choose the level of abstraction, as well as the kind of composition. \textsc{Heraklit} integrates new concepts with tried and tested ones. Such a framework provides the foundation for a comprehensive Systems Mining as the next step after Process Mining.

\end{abstract}

\section {Introduction}

\let\thefootnote\relax\footnotetext{presented at the \textit{16th International Workshop on Engineering Service-Oriented Applications and Cloud Services}, Heraklion, Greece, September 28-30, 2020}

The development of big service-oriented systems is challenging. Traditionally, models have been a central tool for designing such systems. Currently used modelling methods reach their limits and should be replaced by better concepts. The currently prevailing way of developing service-oriented systems is unsatisfactory in many aspects. The development process and its result must be: (a) more manageable for the developer, (b) easy to understand for the user, (c) less error-prone and verifiable, (d) easier to change, faster reachable and cheaper especially for really large systems. These and similar requirements have long been discussed in the relevant literature.

The development of a complex service-oriented system is always preceded by a planning process in which models are used to formulate the structure, function, intended effects etc. of the intended product. In comparison to other engineering disciplines, models are generally not used very often in computer science and business informatics. This is mainly due to the fact that up to now not much benefit can be derived from models. In the practice of system design nowadays mainly diagrams using the Business Process Modeling Notation (BPMN) are propagated for describing the business logic. Such diagrams are limited to the identification of elementary activities and the representation of the control flow. More comprehensive models that take more aspects into account and are more intuitive would be extremely helpful for computer science and business informatics.

We argue for a modelling method whose models are suitable for much more than just the representation of elementary activities and control flows. In particular, a good modelling method should meet the requirements mentioned above. From a more technical point of view, such a method should:

\begin{itemize}

\item support the structuring of a large service-oriented systems into modules;
\item technically simple, but expressive to compose modules to large service-oriented systems;
\item describe the discrete steps in large systems only locally in individual modules;
\item represent modules intended for implementation and modules not intended for implementation integrated with the same concepts;
\item represent data and consider data dependencies in the control flow;
\item abstract from concrete data in order to create instantiations with the same behavior in a schematic way;
\item add under-specified data aspects in the later design process or in the event of changes of the system systematically;
\item describe activities and events at any level of abstraction and hierarchy levels;
\item generate models that are scalable, changeable and expandable;
\item support the proof that a model has desired properties;
\item extend the proven techniques of Data Mining and Process Mining to general Systems Mining.
\end{itemize}

In this paper we propose \textsc{Heraklit} as a modelling method that meets these requirements. It combines proven mathematically based and intuitively easy to understand concepts that are already used for system specification; we recombine them and complement them with concepts for composition and hierarchical refinement of local components, making this technique suitable for modeling large operational systems.

The objective of this paper is to present an overview on \textsc{Heraklit}. Therefore we shortly introduce the central modelling principles in Section 2. Section 3 presents a concise case study using \textsc{Heraklit}. The paper closes with a disucssion of related work (Section 4) and some conclusions (Section 5).

\section {Principles of \normalfont \textsc{Heraklit}}

\subsection{Big systems}
What are the implications of the statement that a system is ``big''? Firstly, some concepts that suit ``small'' systems do not suit large systems. One of the most obvious of these  concepts is the assumption of global states and steps that update global states \cite{reisig2020component_models}. Global states and steps adequately describe, for example, the behaviour of s small digital circuit. To describe the behaviour of stakeholders of a business as a sequence of global steps, is, however, conceptually not adequate. In a big system, e.g. a business, cause and effect of a step are locally confined; and this confinement is essential to understand behaviour. As another specific concept, a big system requires conventions to confine validity of names, i.e.\ to avoid globally valid names, with a few exceptions such as URLs.

In \textsc{Heraklit}, single behaviours (runs, executions) of a subsystem can be represented by means of states and steps that are global only within the subsystem. Upon composing two such systems, those local states and steps are not necessarily embedded into global states and steps of the composed system. Instead, single behaviours of the composed system are represented without assuming global states and steps. Local names of a subsystem are confined to the subsystem and its direct neighboring subsystems.

\subsection {Composition of systems}
Every ``big'' real life system is composed from subsystems that are mutually related: they may exchange messages or jointly execute activities. The composition of subsystems is particularly challenging, as the subsystems themselves may be represented in different languages, modelling methods, etc.\cite{frank2014:modeling} Modelling techniques for such systems must provide means to compose models of subsystems. Many modelling techniques provide such means; they all come with specific, frequently parameterized composition operators, concentrating on special ways to exchange data, e.g.\ synchronously or asynchronously. A ``big'' system, composed from many systems $S_1, \ldots , S_n$, is favorably written 

\begin{equation}
	S = S_1\bullet \dots \bullet S_n,
    \label{equ:composition}
\end{equation}

with ``$\bullet$'' being any version of a composition operator. This bracket free notation requires that the composition operator is associative, i.e.\ that for any three models $R$, $S$, $T$ hold: $(R \bullet S) \bullet T = R \bullet (S \bullet T)$. Typical examples for the notation \eqref{equ:composition} include supply chains, sequences of production machines in a factory, etc. Associativity of composition is rarely discussed explicitly, but frequently assumed without saying \cite{reisig2019associative}.

\textsc{Heraklit} offers a simple, universally employable and associative composition operator. In \textsc{Heraklit}, the diversity of specific, parameterized composition operators is  expressed by help of \textit{adapters}: Specific aspects and properties of the composition $R \bullet S$ of two models $R$ and $S$ are formulated in an adapter $A$, such that $R \bullet A \bullet S$ expresses the wanted properties. The advantages of this concept are obvious: One technical composition operator fits all content-wise requirements, adapters can themselves be composed, etc.

\subsection{Abstraction and refinement}
A number of general principles has been proposed in literature, to adequately cover the abstraction and refinement of systems.  In particular, it is most useful to start out with an abstract specification and to refine it systematically, such that properties of the refined system imply the relevant properties of the abstract system. Vice versa, a given system may be abstracted, yielding a more compact version.

Abstraction and refinement should harmonize with the composition. To refine a part $T$ of a system $S$, one would partition $S$ into $T$ and the environment of $T$, and then refine $T$. The remaining subsystems in the environment of $T$ should not be affected by this procedure. Systems on different refinement levels should be composable; an overall concept of hierarchy levels for subsystems should not be required.  \textsc{Heraklit} suggests concepts for refinement and abstraction that respect these requirements.

\subsection{Modelling of data and things equally}
In a big system, data, physical items, algorithms, activities of persons, steps of organizations, etc.,  are  entangled. They must be modelled by similar means that differentiate between them only in pragmatical aspects: data can be generated, deleted, transformed into different representations, manipulated by computers, copied, updated, composed, etc. Physical items behave differently: A physical item always occupies a distinguished place in space. In models, one frequently does not want to distinguish 
``equal'' items explicitly; their number matters.

\subsection{Behaviour}
The behaviour of a large system is composed of single actions. An action updates some local state components. It is up to the modeler to embed local state components into more global views, if wanted. For a really large system, a single execution (run) should not be represented as a sequence of actions (though one may argue that all behaviour occurs along a global time scale). Independence of actions should explicitly be represented and not be spoiled by representing them in an arbitrary order.

\textsc{Heraklit} suggests to base the description of behaviour on Petri nets with data carrying tokens \cite{reisig2013understanding}. 
This choice is motivated by multiple aspects:
\begin{itemize}
\item Petri nets can easily be specialized to include interfaces: Just select some places, transitions, and even arcs to serve as interface elements.
\item The composition of Petri nets with interfaces is again a Petri net with interfaces.
\item Petri nets suggest the notion of concurrent runs that partially order actions of a run, thus, avoiding them to be mapped onto a global time scale.
\end{itemize}

\subsection{Describing systems on a schematic level} 
Data, real life items, as well as entire systems must be describable on an abstract, schematic level. In particular, it must be possible to describe just the existence of data, items, functions, etc., without any concrete description of how they look like, how many of them there are, etc. On this schematic level, it should be possible to describe activities in systems, e.g.\ the principles of executing a client’s order of an enterprise. A concrete enterprise is then an instantiation of the schema.

\textsc{Heraklit} provides techniques to model such schemata, and to characterize concrete enterprises as instantiations of such a schema. Here, we adapt notions such as structures, signatures and instantiations of signatures, that are well-known from first order logic and algebraic specifications. (Technically, a signature is just a set of sorted symbols for sets, constants, and functions. An instantiation interprets these symbols consistently). We extend signatures by requirements to exclude ``unwanted'' instantiations, in the spirit of specification languages such as the Z language. 

Signatures and their instantiations can naturally be transferred to define Petri net schemata – we call them \textsc{Heraklit} schemata. Such a schema can be instantiated in different ways; each instantiation results in a concrete Petri net. This concept is useful to model, for example, not just a distinguished business, but a class of businesses that all follow the same business rules. Hence, \textsc{Heraklit} strongly supports the idea of reference modelling, a core topic of business informatics \cite{rehse2019:situation_ref_mod}. 

\subsection{Verification}
The notion of correctness has many implications for big systems. Some ideal properties of a big system can be composed of corresponding properties of the component systems. Not all relevant properties can formally be captured, yet they deserve a proper framework to reason about them. Particularly interesting are methods to prove properties at run-time.

\textsc{Heraklit} integrates a number of formal and semi-formal verification techniques to support structured arguments about the correct behaviour of modules.

\section{Modules and their composition}

\subsection{Modules}
In Section 1 we discussed a number of principles that are inevitable for modelling big systems: no globally effective structures, associative composition of models of any two systems, composition must be compatible with abstraction, modelling of data and real items, modelling of behaviour, parameterized models. Now we must model systems in such a way that all these principles are met.   

We start out with the obvious observation that a real system in general consists of interdependent subsystems. This paves the way for the central notion of \textsc{Heraklit}-modules: A \textsc{Heraklit} module is a model, graphically depicted as a rectangle, with two decisive components:
\begin{itemize}

\item Its inner: this may be any kind of graph or text. Three variants are frequent: (a) the inner consists only of the name of the module, (b) it consists of (connected) submodules, (c) it describes dynamic behaviour.

\item Its surface: this consists of gates, each gate is labelled, i.e.\ inscribed by a symbol. The gates of the surface are arranged on the surface of the module’s rectangle. Alternatively, each gate is represented as a line, linking the module’s rectangle with the gate’s label.
\end{itemize}

The following Fig.~\ref{fig:modules} shows typical \textsc{Heraklit} modules.

\subsection{Composition of modules}
Composing two modules $A$ and $B$ follows a simple idea: two equally labelled gates of $A$ and $B$ are “glued” and turned into an inner element of the module $A\bullet B$. However, in this simple version, the composition is fundamentally flawed: Upon composing three or more modules, the order of composition matters: for three modules $A$, $B$, and $C$, the two modules $(A \bullet B) \bullet C$ and $A \bullet (B \bullet C)$ differ from one another. In technical terms: this version of composition is not associative. But associativity is a central requirement, as discussed in Chapter 1.2.

To solve this problem, we return to modules shaped $S = S_1 \bullet \dots  \bullet S_n$. As discussed in Sec. 1.2: each module $S_i$ generally has a left and a right neighbor ($S_0$ has no left, $S_n$ has no right neighbor). $S$ is composed by composing $S_{i-1}$ with $S_i$ (for $i = 2, \dots,n$). In the real world, systems frequently exhibit 
this kind of structure, physically or conceptually. 

Therefore, \textsc{Heraklit} partitions the surface of a module $L$ into its left and right interface, written $^\ast L$ and $L ^\ast$, resp. To compose two modules $L$ and $M$, equally labelled gates of $L ^\ast$ and $^\ast M$ are glued and turn into inner elements of $L \bullet M$. The remaining elements of $L ^\ast $ go to $(L  \bullet M) ^\ast$ (together with $M^ \ast $), and the remaining elements of $^\ast M$ go to $^\ast (L \bullet M)$ (together with $ ^\ast L$). Most important: A general theorem guarantees that this kind of composition is associative \cite{reisig2019associative}.

\section{Case study: a service system}

\subsection{The different modules of the system}
Today, many organizations offer a complex service portfolio for their customers or clients \cite{boehmann2014service,chesbrough2006manifesto}. Typical examples are banking or financial services, insurance services, legal services, and the medical or health services offered by a hospital or a medical center.

\begin{figure}[htb]
    \centering
    \includegraphics[width=0.66\linewidth]{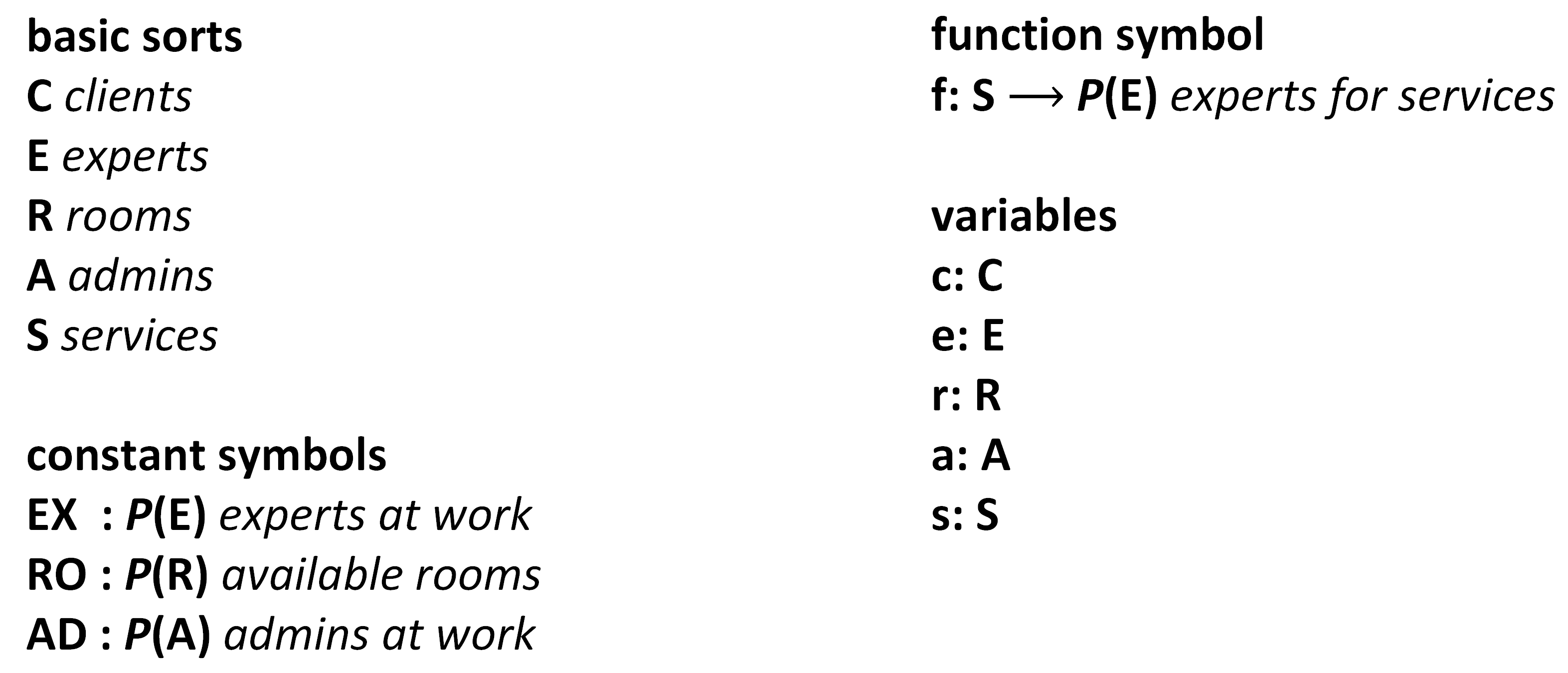}
    \caption{Signature of the service system}
    \label{fig:signature}
\end{figure}

Here, we model the organization of such a service system, serving clients, customers, or patients that want confidential consultation about particular services or a particular treatment, provided by experts. 

Fig.~\ref{fig:signature} shows the signature of the system: there are five sorts of elements in a service system, indicated by $C$, $E$, $R$, $A$, and $S$. Their intuitive meaning is indicated in italic. In a concrete service system, there are sets of experts, available consulting rooms, and admins, symbolically represented by $EX$, $RO$ and $AD$. Their type is $P(E)$, $P(R)$, and $P(A)$, resp., with $P(\cdot)$ standing for 
``powerset''. Furthermore, we need a function symbol $f$ and five variables, one for each basic sort. An \textit{instantiation} assigns each basic sort an arbitrarily chosen concrete set, each constant symbol a set of elements of the indicated sort, and $f$ a function that assigns each service the set of experts that offer consultations for this service.

\begin{figure}
      \centering
      \subcaptionbox{clients\label{fig:clients}}
        {\includegraphics[width=0.50\textwidth]{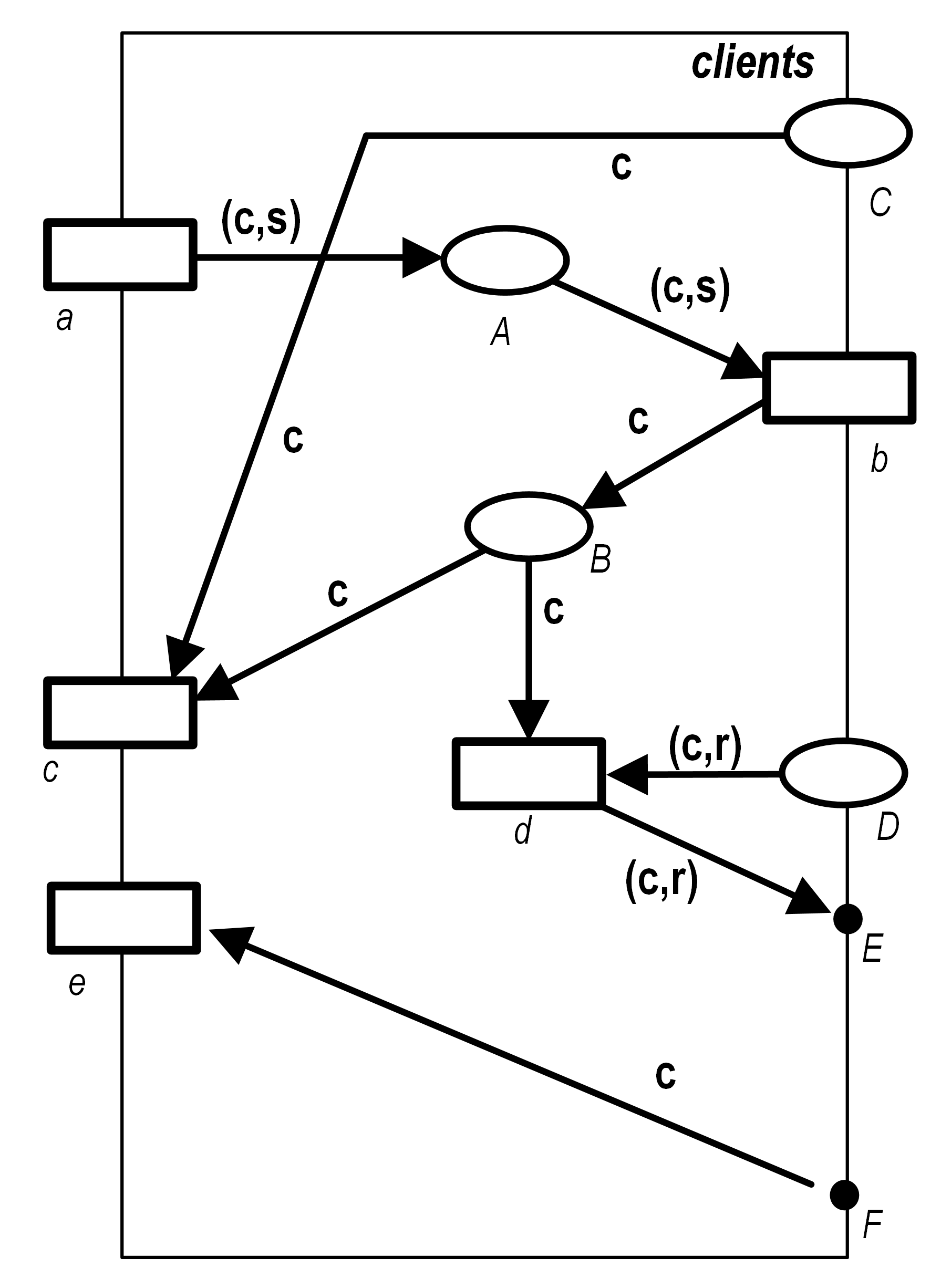}}
      \subcaptionbox{experts\label{fig:experts}}
        {\includegraphics[width=0.375\textwidth]{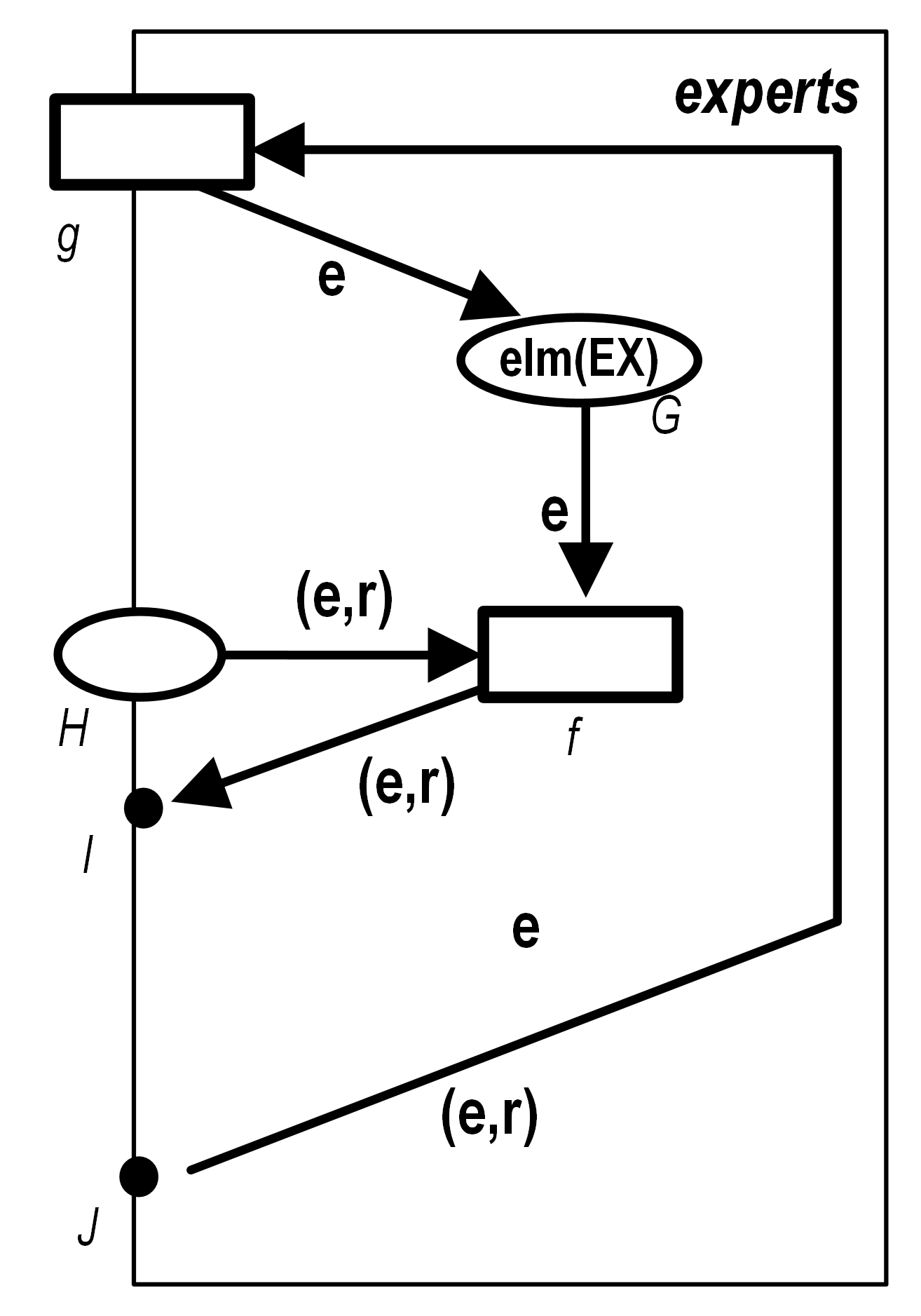}}

      \subcaptionbox{admin\label{fig:admin}}
        {\includegraphics[width=0.75\textwidth]{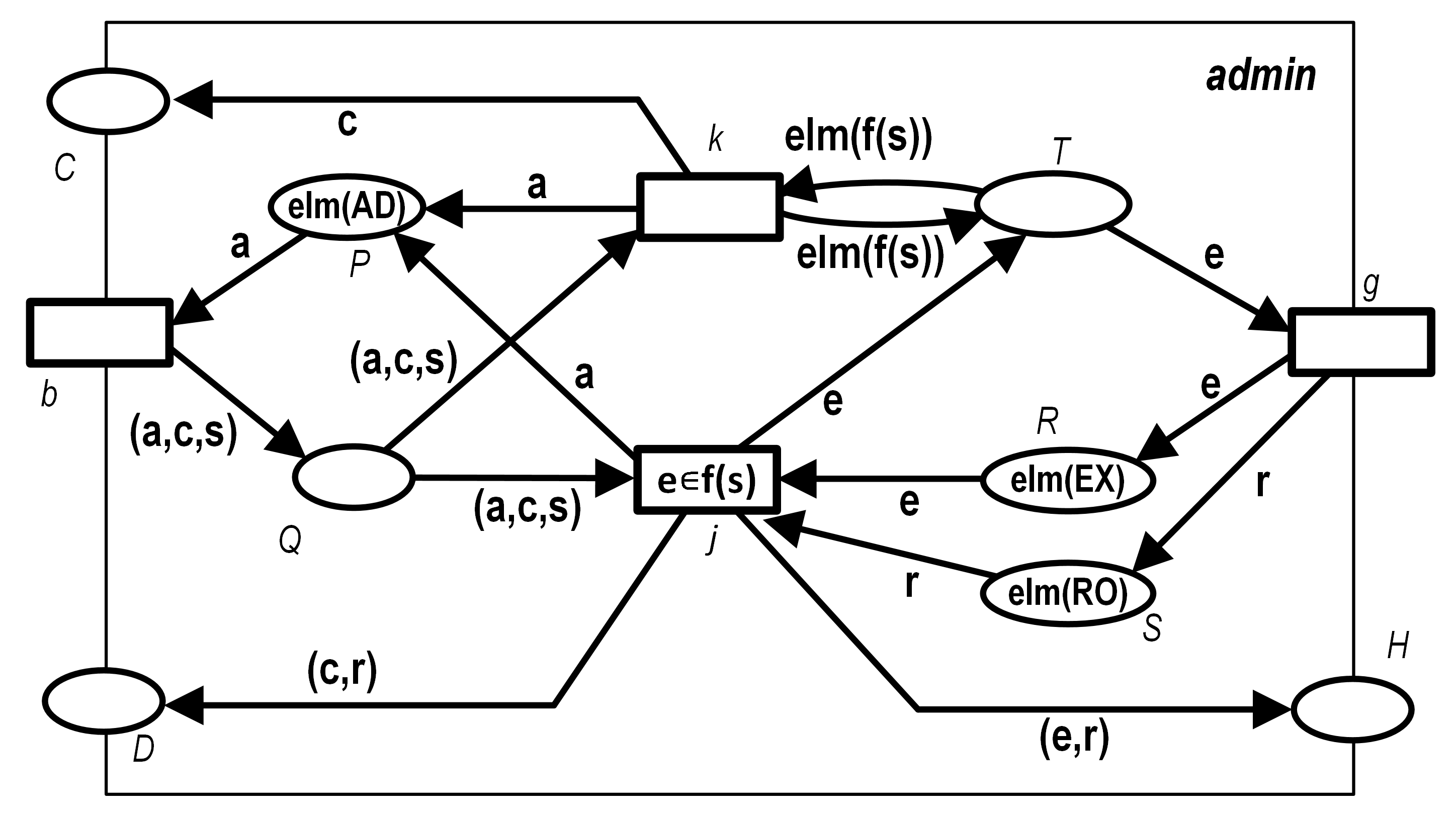}}
        
      \subcaptionbox{consulting rooms\label{fig:rooms}}
        {\includegraphics[width=0.75\textwidth]{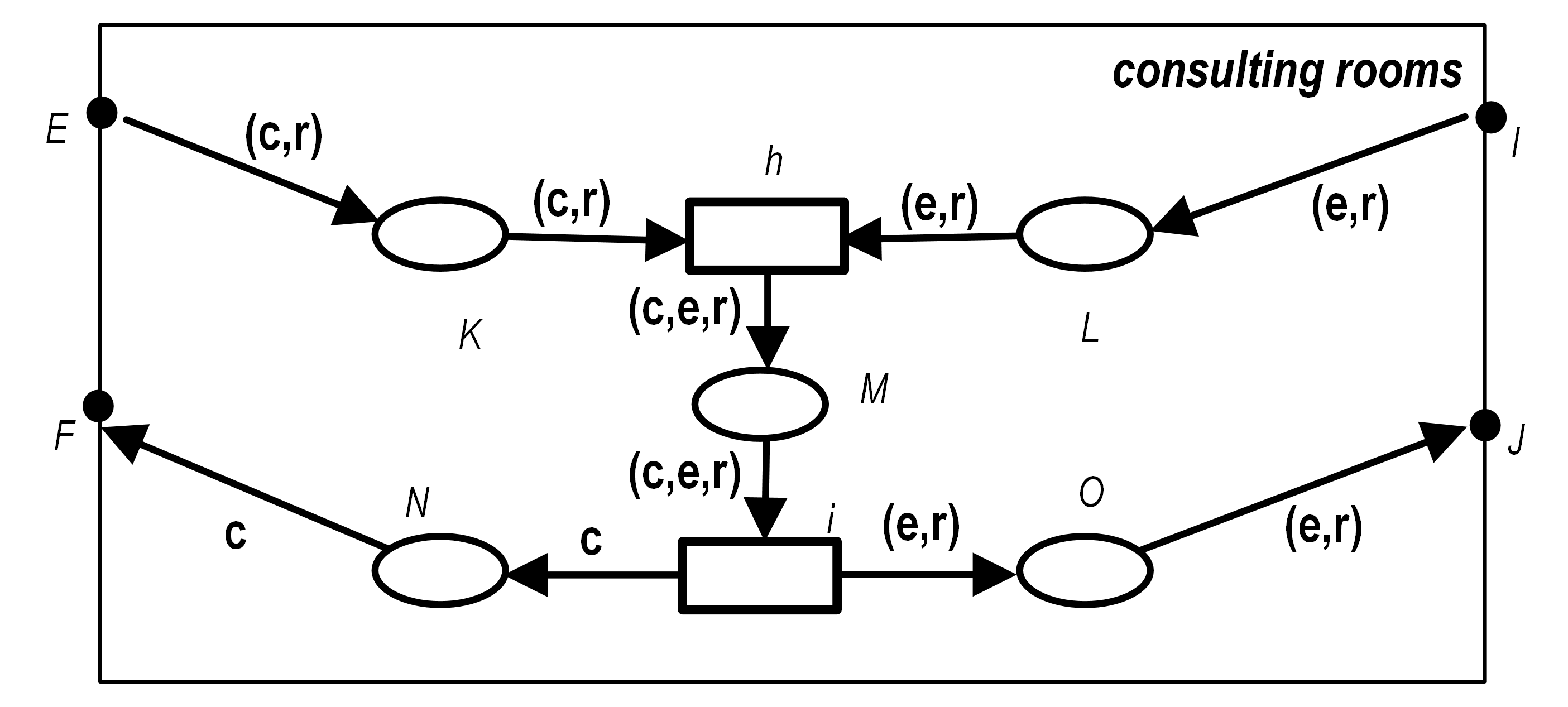}}

      \caption{The four modules of the system}\label{fig:modules}
    \end{figure}

Fig.~\ref{fig:clients} shows a module that represents the behaviour of clients: For every instantiation of the variables $c$ and $s$ by a client and a service, resp., transition $a$ is enabled. Transition $a$ represents the policy that any client may enter the service system with any kind of wish for consultation for a service $s$. Hence, place $A$ may eventually hold any number of tokens, with each token consisting of a client and a service. Transition $b$ indicates the service systems’s help desk, accepting each client’s wishes and asking them to wait at place $B$. There, a client will eventually receive a message either at place $C$ or at place $D$. A message at place $C$ indicates that no expert is available; so the client leaves the service system along transition $c$. A message at place $D$ indicates that the client should proceed to the consulting room named or numbered $r$. The client will do so along transition $d$ and arrow $E$. He will later on return along arrow $F$ and leave the service system by transition $e$.

The module in Fig.~\ref{fig:experts} represents the behaviour of the service system’s experts. There is a set of experts, depicted as $EX$, fixed when the schema is instantiated, and initially represented as unengaged at place $G$. One might expect this to be expressed by the symbol $EX$ at place $G$. However, this would indicate one token at place $G$. This is not what we want: we want each single expert to be represented as a token. This is achieved by means of the function $elm$: Applied to a token that represents a set $M$, $elm(M)$ returns each element of $M$ as a token. For an expert $e$, the message $(e,r)$ arriving at place $H$ indicates that $e$ must go to consulting room $r$, due to transition $f$ and arc $I$. He will eventually return along arc $J$, release room $r$, and will be again unengaged at place $G$.

The module in Fig.~\ref{fig:rooms} shows the consulting rooms: A client $c$ and an expert $e$ arriving at room $r$ along the arcs $E$ and $I$, resp., start their consultation by transition $h$, end it by $i$, and leave the room by arcs $F$ and $J$.



\begin{figure}[H]
    \centering
    \includegraphics[width=1\textheight, angle=90]{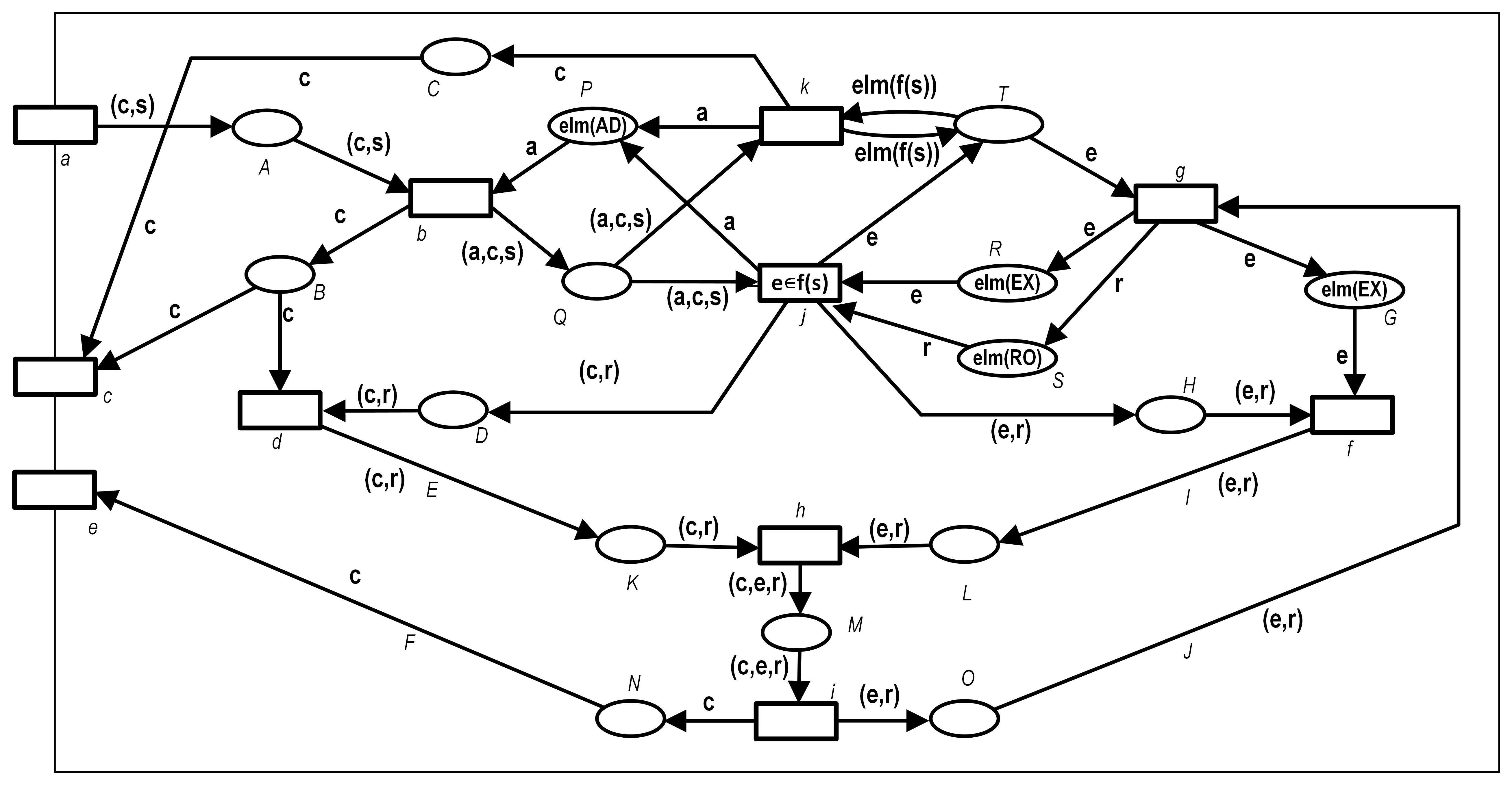}
    \caption{Overall model of a service system}
    \label{fig:overall}
\end{figure}



The behaviour of clients, experts, and the consulting rooms must be properly synchronized. The admin module of Fig.~\ref{fig:admin} organizes this. Place $P$ initially contains each admin as a token (we employ again the function $elm$ as explained above for the experts). An admin $a$ engages with a client $c$ and their request for an expert for service $s$, along transition $b$. A token $(a,c,s)$ on place $Q$ then continues either along transition $k$ or transition $j$. Transition $j$ requires an expert $e$ on place $R$, such that $e$ offers the service $s$. The inscription of $j$ indicates this requirement. $R$ always contains a “digital twin” for each expert that is not engaged with a client. The place $S$ always contains a digital twin of each empty consulting room. Hence, transition $j$ is enabled with proper instantiations of all five variables $a,c,s,e,$ and $r$. The occurrence of $j$ then renders the admin $a$ available in $P$ for new clients, sends messages to the client $c$, and the expert $e$ to proceed to room $r$, and moves the digital twin of $e$ to place $T$. This way, the digital twin of each expert $e$ is either a token in $R$ or in $T$. With $e$ in $T$, the expert $e$ eventually indicates by transition $g$ that they finished their consultation and they release the room $r$. Finally, transition $k$ manages the case where for a token $(a,c,s)$ no expert for service $s$ is available in $R$. As discussed above, the digital twin of each such expert is a token in $T$. Hence, all tokens in the set $f(s)$ of experts for $s$ are in $T$. This is “tested” by means of the loop between $k$ and $T$. Occurrence of $k$ then renders the admin $a$ available in $P$ for new clients and sends a corresponding message to the client $c$.
Notice the subtle treatment of experts and rooms as a scarce resource: If no corresponding expert is available, a client is turned away, as it may take too long until an expert for $s$ is available. But if no room is available, the client is just waiting as long as one room will be available. 


\subsection{Overall model and abstract composition}

Fig.~\ref{fig:overall} finally “glues” the four modules into one big module. In \textsc{Heraklit}, this can just be written as: $    clients \bullet admin \bullet consulting \ rooms \bullet experts.
$

\begin{figure}[htb]
    \centering
    \includegraphics[width=0.6\textwidth]{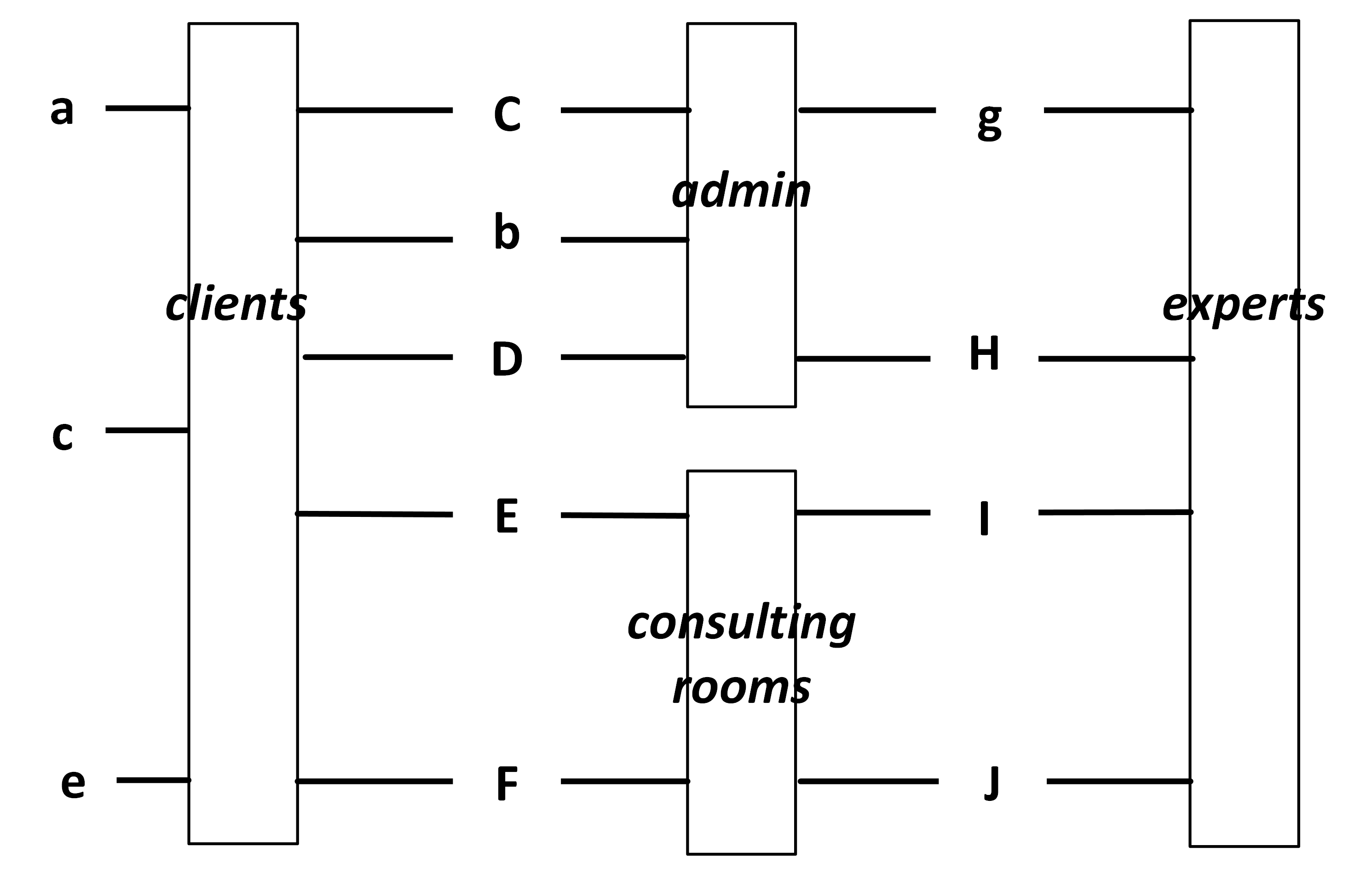}
    \caption{Abstract composition of the overall model}
    \label{fig:abstract_composition}
\end{figure}

Similarly, it is possible to construct an abstract composition of the system. Fig.~\ref{fig:abstract_composition} depicts such a composition of the four abstract modules by using the abstraction operator $[\cdot]$, which deletes the inner structure of a module. Formally written as: $    [clients] \bullet [admin] \bullet [consulting \ rooms] \bullet [experts].
$

\section{Related work}

Modelling is typically understood as an interdisciplinary field that is used in many different disciplines as a method or instrument to capture knowledge or to assist other (research) actions \cite{frank2014:modeling,beverungen2020bpm}. As we discuss above, \textsc{Heraklit} mainly does not invent new modeling concepts but integrates proven and well-known modelling approaches. Compared to other integrated approaches which currently dominate the modelling practice, e.g. BPMN, \textsc{Heraklit} provides integrated means to descrive model structure, data, and behaviour.  In the central concept of a module, \textsc{Heraklit} combines three proven, intuitively easy to understand, and mathematically sound concepts that have been used for the specification of systems in the past:

1. Abstract data types and algebraic specifications for the formulation of concrete and abstract data: since the 1970s such specifications have been used, built into specification languages, and often used for (domain-specific) modelling. The book \cite{sanella20212algebraic} presents systematically the theoretical foundations and some applications of algebraic specifications. Abstract state machines \cite{gurevich1993algebras} also belong to this context.

2. Petri nets for formulating dynamic behaviour: \textsc{Heraklit} uses the central ideas of Petri nets. A step of a system, especially a large system, has locally limited causes and effects.  This allows processes to be described without having to use global states and globally effective steps. This concept from the early 1960s \cite{petri1962diss} was generalized at the beginning of the 1980s with predicate logic and \textit{colored marks} \cite{genrich1979predicatenets,jensen1982colourednets}. The connection with algebraic specifications is established by \cite{reisig1991algebraic}. \textsc{Heraklit} adds two decisive aspects to this view: uninterpreted constant symbols for sets in places that use the $elm$ function to hold instantiations with many possible initial marks, and the $elm$ function as an inscription for an arrow to describe flexible mark flow.

3. The composition calculus for structuring large systems: this calculus with its widely applicable associative composition operator is the most recent contribution to the foundations of \textsc{Heraklit}. The obvious idea, often discussed in the literature, of modeling composition as a fusion of the interface elements of modules is supplemented by the distinction of left and right interface elements, and composition $A \bullet B$ as a fusion of right interface elements of $A$ with left interface elements of $B$. According to \cite{reisig2019associative,reisig2020component_models}, this composition is associative (as opposed to the naive fusion of interface elements); it also has a number of other useful properties. In particular, this composition is compatible with refinement/coarseness and with individual (distributed) runs. 

These three theoretical principles harmonize with each other and generate further \textit{best practice} concepts that contribute to a methodical approach to modeling with \textsc{Heraklit}, and which will only be touched upon in this paper. On the down-side, industrially mature modelling tools for \textsc{Heraklit} are still under development.

\section{Conclusions}
The presented case study clearly demonstrates how \textsc{Heraklit} provides an integrated view on structure, data, and locally defined behaviour. Hence, \textsc{Heraklit} covers all central aspects of every computer-integrated system. Such a description can be used for different purposes, e.g.\ business process management, service engineering, software analysis, design, verification, and development. The used techniques are well-known but combined in a novel and innovative way.

By providing such an integrated method for system specification, \textsc{Heraklit} paves the way for many important innovations which are currently so much in need \cite{beverungen2020bpm,houy2010:bpm_large}. In particular, we like to introduce the idea of \textit{Systems Mining}.
While Data Mining and Process Mining \cite{aalst2012process_mining} exploit the knowledge implicitly represented in data tuples and event sequences, respectively, Systems Mining is able to analyze the structure, data, and behaviour of a system. For such analysis, \textsc{Heraklit} provides the necessary techniques to specify all essential characteristics of a system. The observed structure of the system can be represented as modules, the observed data is captured by both concrete and abstract data structures, and the observed behaviour is specified as (distributed) runs.  Based on such a powerful framework, Systems Mining provide a much richer picture of and deeper insights into big systems. 

The presented case study of a service system illustrates powerful possibilities. Based on these \textsc{Heraklit} models, Systems Mining can answer a wide spectrum of interesting questions: (1) Do typical communications patterns between the modules of the system exist? (2) Which services are often requested by customers? (3) Do customers follow particular patterns for requesting services? (4) Which particular service requests and assignments of experts and rooms typically cause long waiting times for a customer? (5) Are there particular behaviour patterns and service requests which typically cause customers to leave the service system without getting a service or treatment?

Such questions and many more can easily be specified with \textsc{Heraklit}. Additionally, \textsc{Heraklit} provides a richer foundation for predictive and prescriptive process management as well as deeper insights for explaining process behaviour \cite{evermann2017prediction,mehdiyev2020explanation}.  Hence, \textsc{Heraklit} lays the foundation for the next step after Process Mining.
%
%
%

\bibliographystyle{splncs04}
\bibliography{main}





\end{document}